\documentclass{PoS}

\title{Cherenkov Telescope Array Data Management}

\ShortTitle{CTA Data Management}

\author{\speaker{G. Lamanna}$^{a}$, L.A. Antonelli$^{b,c}$, J.L. Contreras$^d$, J. Kn\"odlseder$^e$, K. Kosack$^f$, N. Neyroud$^{a}$, A. Aboudan$^g$, L. Arrabito$^h$, C. Barbier$^{a}$, D. Bastieri$^i$, C. Boisson$^{j}$, S. Brau-Nogu{\'e}$^e$, J. Bregeon$^h$, A. Bulgarelli$^g$,  A. Carosi$^{b,c}$, A. Costa$^{k}$, G. De Cesare$^g$, R. de los Reyes$^l$, V. Fioretti$^g$, S. Gallozzi$^b$, J. Jacquemier$^{a}$, B. Khelifi$^m$, J. Kocot$^{n}$, S. Lombardi$^{b,c}$, F. Lucarelli$^{b,c}$, E. Lyard$^{o}$, G. Maier$^{p}$, P. Massimino$^k$, J. P. Osborne$^q$, M. Perri$^{b,c}$, J. Rico$^{r}$, D. A. Sanchez$^{a}$, K. Satalecka$^d$, H. Siejkowski$^{n}$, T. Stolarczyk$^{f}$, T. Szepieniec$^{n}$, V. Testa$^b$, R. Walter$^{o}$, J.E. Ward$^{r}$ and A. Zoli$^g$ for the CTA consortium\thanks{Full consortium list at http://cta-observatory.org}\\ 
\llap{$^a$}LAPP, Universit{\'e} Savoie Mont-Blanc, CNRS/IN2P3, F-74941 Annecy-le-Vieux, France\\
\llap{$^b$}INAF - Osservatorio Astronomico di Roma, Via Frascati, 33, I-00040 Monte Porzio Catone\\
\llap{$^c$}ASI Science Data Center, Via del Politecnico, s.n.c., I-00133 Rome, Italy\\
\llap{$^d$}Grupo de Altas Energ\'ias, Universidad Complutense de Madrid, Spain \\
\llap{$^e$}IRAP, 9, avenue du Colonel Roche BP 44346 - 31028 Toulouse, France\\
\llap{$^f$}CEA Saclay - Irfu/SAp - Orme des merisiers, Bât. 709 - 91191 Gif-sur-Yvette Cedex, France\\
\llap{$^g$}INAF/IASF, Via Piero Gobetti 101,  40129 Bologna, Italy\\
\llap{$^h$}LUPM, CNRS-IN2P3 Place Eug{\'e}ne Bataillon - CC 72 34095 Montpellier, France\\ 
\llap{$^i$}Dipartimento di Fisica, Universita degli studi di Padova, Italy\\
\llap{$^j$}Observatoire de Paris, LUTH, CNRS, Universit{\'e} Paris Diderot, Paris, France\\
\llap{$^k$}INAF - Osservatorio Astrofisico di Catania, Via S.Sofia 78, 95123 Catania, Italy\\
\llap{$^l$}Max-Planck-Institut f\"ur Kernphysik Saupfercheckweg 1, 69117 Heidelberg, Germany\\
\llap{$^m$}APC, IN2P3/CNRS, Universit{\'e} Paris-Diderot, Paris, France\\
\llap{$^n$}Academic Computer Centre CYFRONET AGH, Krak{\'o}w, Poland\\
\llap{$^o$}ISDC, University of Geneva, Switzerland\\
\llap{$^p$}Deutsches Elektronen-Synchrotron  Platanenallee 6, Zeuthen, Germany\\
\llap{$^q$}Dept. of Physics \& Astronomy, University of Leicester, Leicester LE1 7RH, GB\\ 
\llap{$^r$}IFAE, Edifici Cn. Universitat Autonoma de Barcelona 08193 Bellaterra (Barcelona) Spain\\
E-mail: \email{giovanni.lamanna@lapp.in2p3.fr}
}

\abstract{Very High Energy gamma-ray astronomy with the Cherenkov Telescope Array (CTA) is evolving towards the model of a public observatory. Handling, processing and archiving the large amount of data generated by the CTA instruments and delivering scientific products are some of the challenges in designing the CTA Data Management. The participation of scientists from within CTA Consortium and from the greater worldwide scientific community necessitates a sophisticated scientific analysis system capable of providing unified and efficient user access to data, software and computing resources. Data Management is designed to respond to three main issues: (i) the treatment and flow of data from remote telescopes; (ii) "big-data" archiving and processing; (iii) and open data access. In this communication the overall technical design of the CTA Data Management,  current major developments and prototypes are presented.}

\FullConference{The 34th International Cosmic Ray Conference,\\
		30 July- 6 August, 2015\\
		The Hague, The Netherlands}

\begin{document}

\section{Overall Concept}
CTA is a new observatory for very high-energy (VHE) gamma rays~\cite{cta}.
The Data Management (DATA) project of CTA concerns all major components for both data-flow administration and the
scientific data production and analysis for CTA. The main scope of the project is the design of the
CTA Science Data Centre (SDC), which is in charge of the off-site handling of
data reduction, Monte Carlo (MC) simulations, data archiving and data
dissemination. The remote (e.g. intercontinental) transmission of data
from CTA sites to the CTA archive is one of the key services that the
SDC administers at both ends: off and on the CTA site.
The development and provision of software and middle-ware services for
dissemination including observation proposal handling is a task that
DATA guarantees to be interfaced with the Operation Centre.
The services and components that DATA is in charge
of at the CTA sites (Telescope Array Control Centres) include: the execution of on-site scientific data
reduction pipelines, the real-time analysis, the on-site
temporary archive system as well as the data quality monitoring. 

\section{Summary of Design}
\begin{figure*}[!t]
 \center
 \includegraphics[width=12cm]{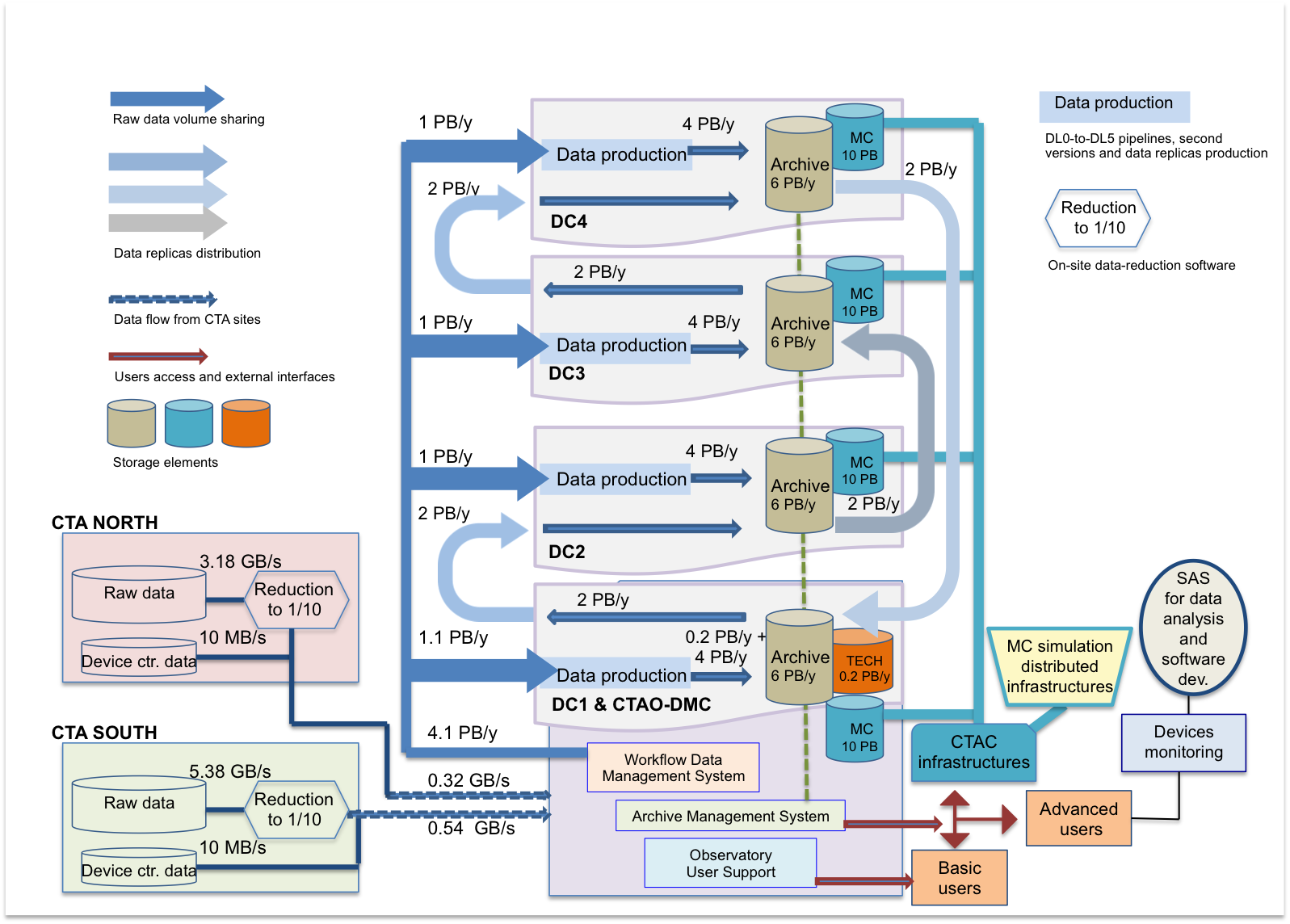}
 \caption{CTAO data volume management. Raw data and device control data are transferred from CTA sites to the CTAO Data Management Centre along the network. Data are then distributed over four data-centres participating in the processing and archive of data. The total data volume, including replicas, is managed by the four data-centres sharing the CTAO storage. }
 \label{fig:introflow}
\end{figure*}

The DATA design is inspired by the lessons learned from current and past Imaging Atmospheric Cherenkov Telescopes, from CTA telescopes prototype, from existing astronomical observatories, and finally from the technical know-how of major computing / data centres and e-infrastructures that serve large international projects. Figure~\ref{fig:introflow} depicts the main path and rate of data within the CTA Observatory (CTAO). On each CTA site, the data rates are based on the event rates from Cherenkov and night-sky-background triggers
registered by the telescope array.
The rates depend strongly on: the number and type of telescopes in the
two arrays ($\sim$130 in the current assumptions), the number of pixels per camera, the nominal trigger
rates, the length (in time) of the pixel readout windows, the number of
samples per unit time, and the number of bytes recorded per sample. 
Some pre-processing and filtering of stereoscopic Cherenkov events
will affect the nominal data rates, which will result in 5.4 GB/s for
CTA south (with $\sim$100 telescopes) and 3.2 GB/s from CTA north (with $\sim$ 30 telescopes). They also include 20\% calibration data and 10 MB/s of device monitoring and control data for each site, for a resulting total data rate of about 8.6 GB/s.
The remote connection to the CTA site candidates must satisfy the bandwidth
capacity of 1 Gb/s, which makes the issue of the exported data size critical.  
During the construction phase, DATA will
develop a data volume reduction system that cuts the nominal data rate by a
factor of 10, thus implying an output rate of
only 0.32 GB/s and 0.54 GB/s data from CTA north and south
respectively. These data rates are valid over an annual duty cycle of 1314 h and they will correspond to an equivalent and continuous data volume flux of 0.38 Gb/s and 0.65 Gb/s, well
manageable (with respect to the latency requirements) with
a local temporary storage unit on the CTA sites, and a 1 Gb/s effective
network for off-site data export. Current perspectives for the
availability of 10 Gb/s network from candidate CTA sites
exist.

The exported data are received by the \emph{ingestion unit} of
the CTA Archive system, which is operated (together with all main
work-flow management services) by a dedicated Observatory Data
Management Centre (CTAO-DMC). The proposed CTAO computing model is
built upon a Distributed Computing Infrastructure (DCI) approach, in
which a limited number of first-class data-and-computing
centres share the workload of archiving and processing the CTA
data. The baseline of DCI model adopted to estimate the work and
investment distribution is made of four centres (from DC1 to DC4),
equally sharing the CTAO data workload (Figure~\ref{fig:introflow}).
One out of these four centres also hosts the CTAO-DMC. The CTAO-DMC plus the four DCs correspond together to the proposed implementation of the CTA \emph{Science Data Centre}. The CTAO-DMC simultaneously guarantees: (i) the
orchestration of the relay-mode activities among the four centres,
while centrally managing the database of the CTAO archive; (ii) all
interface services with the users, providing tools to the Observatory User Support group and to advanced users for archive
queries and data processing or to access technical data for devices
monitoring purposes as well as User Support; (iii) the integration of the Scientific Analysis System (SAS) with the CTAO and the
CTA Consortium (CTAC) infrastructures, which are also DCIs and are used currently
for MC simulation production (through the existing CTA Computing Grid infrastructure - CTACG). Each data-centre will permanently archive one half of the
full CTA data: it receives 25\% of raw data to be
processed and will play the role of back-up centre archiving the
replica of another 25\% of the data, coming form another centre. The
geographical location of these centres is not specified, and from the technical point of view any place where a qualified (such as WLCG Tier1-like) data centre is willing to contribute to the implementation of the
model (including in the CTA array host countries) may be considered. The total volume
to be managed by the CTAO Archive is of the order of 27 PB/year, when
all data-set versions and backup replicas are considered. This will
correspond to a permanent archive of the order of more than 400 PB in 2031. 

The computing needs are less critical: peak values of less
than 10$^4$ CPU cores are expected for the annual data
processing. The CTA data processing is based on several levels of processing: the
MC simulation pipeline, the low-level data processing
pipelines, the high-level science software tools and the Virtual
Observatory (VO) data access services. The speed and performance requirements of the design of the data
processing pipelines necessitate a high level of parallelization. This
will imply a large effort to reducing the I/O speed and CPU time
required. This is particularly relevant in the case of real-time or
on-site analysis, which initially will use similar software to that
used off-site, but with more strict requirements on speed and looser
requirements on accuracy. However, the adoption of new computing
architectures (e.g. GPGPU, ARM processor) to reduce the cost and
improve the performance of the data streaming and accelerate the
processing, are not excluded in the current design.

MC simulations are required to characterize the performance
and response of the instrument to Cherenkov light emitted in extensive
air showers. Simulations are also necessary for the development of
reconstruction and analysis tools. Therefore a large quantity of
simulated data (20 PB) will be produced prior to the operations phase (during the construction phase) and permanently archived. MC processing
is also envisioned during the operation phase to produce up-to-date
Instrument Response Functions (IRF), and for validation of new
algorithms and software versions. The format of simulated data and
software for reconstruction and analysis are identical to those
applied to observation data (with some extra information attached,
corresponding to simulated parameters).

Off-site, in each DC$_i$ data center, the data production consists 
of a series of processing steps that transform archived
(reduced) raw data DL0 (Data Level 0) to calibrated camera data (DL1),
then to reconstructed shower parameters such as energy, direction, and
particle ID (DL2), and finally to  high-level observatory products
comprised of selected gamma-like events, instrument response tables, and
housekeeping data (DL3).  DL3 data will have a total volume of about 2\% of the DL0 data
volume and guaranteed access will be provided in the CTA archive to
basic users  (e.g. Guest Observers and Archive Users).  The science tools are then used either automatically or
by users to produce DL4 (e.g. spectra, sky-maps).  Finally (DL5) legacy
observatory data, such as CTA survey sky maps or the CTA source
catalog will be produced. When all
replicas and versions are considered, according to user analysis requirements
and archive optimization, the data processing and archiving processes generate 27 PB/year
data out of incoming reduced DL0 4 PB data per year
(Figure~\ref{fig:introflow}). The CTAO
computing model makes use of disk ($\sim$20\%) and tape storage ($\sim$80\%) to guarantee
effective storage, access and throughput of all data, while trying to
reduce the associated costs. 

Another big challenge for the management of data by the CTAO is the
open access to the CTA data. To operate as an open observatory, a
minimum set of services and tools are needed to support the scientific
users of CTA. These services are intended to be mostly web-oriented
and consist of: electronic support services to help Guest Observers in
writing and submitting a proposal to CTA in response to an
Announcement of Opportunity for observing time; user interfaces to
follow the status of an observation, including the scheduling, the
data acquisition, the data processing, the data distribution and the
ingestion of the data in the public archive after the end of the
proprietary period; and finally services for downloading the processed
data (DL3) as well as the software tools that are necessary for
scientific analysis. Web-based information about the data and the
analysis software, including user manuals, cook books, etc. will also
be available. Archive Users will browse the archive to access and
retrieve CTA data of interest selecting events based on specific
criteria (source location, observation time interval, observation
condition criteria, energy range etc.). CTA basic users will analyse
the data will be analysed on their own computing.

DATA will ensure the integration of CTA high-level data (DL4-to-DL5)
within the Virtual Observatory (VO) infrastructure, by adopting and
extending the VO data model standards suitable for the description of
gamma-ray data. High-energy data at this level have never
before been available to the VO community, and this represents a major
step toward unifying the data products from all high-energy
experiments.  Current astronomical metadata standards and VHE
gamma-ray data conventions have been studied for this purpose via a
close working relationship between both VHE and VO scientists. The
extension of standard VO semantic models will allow compatibility with
a large range of existing tools and infrastructures for data
discovery, visualization, and processing. Detailed
\emph{Characterization Data Model} fields will be completed for
high-level products (images, spectra, light curves) and also for event
lists and (Instrument Response Functions) IRFs, allowing scientists
from other backgrounds to discover and manipulate high-level CTA data
products without requiring specialized CTA tools for all
operations. The final goal is to integrate CTAO data in astronomical
multi-wavelength data archives where scientist will be able to combine
them together in a single analysis with data from other
facilities. 

The size and the world-wide scope of the CTA consortium,
along with the desire of CTAC advanced users to access the full
archive and to manage more complex analysis work-flows, demands the
implementation of services to operate a common scientific analysis
platform. In this respect an important baseline of the
IC-infrastructure solution for data access is the \emph{CTA Scientific
  Gateway}: a web-based community-specific set of tools, applications,
and data collections that are integrated together via a web portal,
providing access to resources and services from a distributed
computing infrastructure. The Gateway aims at supporting work-flow
handling, virtualization of hardware, visualization as well as
resource discovery, job execution, access to data collections, and
applications and tools for data analysis. Furthermore the Gateway may
even potentially host all monitoring services of data operation as
well as some remote control or monitoring applications for instruments
and devices when applicable.  The continuous and cooperative software
development within the CTAC requires some consortium shared services
such as a software repository, development tools, version track
services and software validation test benches. Most are currently
implemented and already in use in CTA, and will evolve in the future
into a single web-oriented global platform of services.

Access to the development services, Gateway, and other CTA web
resources will be based on each user's profile and category
(e.g. basic, advanced users, managers, collaboration users, etc). For
such a purpose an \emph{Authentication and Authorization
  infrastructure} is under development and will be applied to extend
the use of the CTA Gateway to any user tuned
according to their own role and/or access rights.

\section{Main developments and prototyping}
Currently the CTAC organizes the DATA development activities around
five main basic components: (i) data model, (ii) archives, (iii)
pipelines, (iv) observer access and (v) IC-infrastructures.

\subsection{Data Model}
In the context of the development of a global Data Model for CTA, some
prototyping work has been conducted around the application of
Compressed FITs format (CFITS) as a file format for Low
Level Data, resulting in a viable and cost-effective solution. For
higher level camera data a flexible format is looked for. Three
formats are under consideration and prototyped: one based on the
\textit{Google Protocol Buffers} specification, a specific
\textit{eventio} format developed for the HESS experiment, and a
\textit{stream} data format based on satellite missions:
packetLib. Packetlib, is a software library that manages complex data
layouts described with XML files, providing introspection, having a
strong memory management and performing a non linear decoding: this,
coupled with the Consultative Committee for Space Data Systems (CCSDS)
space packets standard enables a fast and on-the-fly data
identification, access and routing. For the high level data
comparative studies are conducted among FITS, ROOT and HDF5 file
formats. 
\subsection{Pipelines} $Pipelines$ refers to all software
components necessary to process real and simulated raw data and
produce the final end products needed for science analysis, along with
any associated quality monitoring and technical data. A pipeline is
defined as a sequence of data processing steps that are applied to
data to achieve a high-level goal. The logical design of data
reconstruction and analysis pipelines is well understood, it relies on
the software in use in current experiments such as HESS, MAGIC and
VERITAS, complemented by recent explorative developments from
precursors such as FACT and ASTRI. Main prototyping activities were
around the evaluation of software frameworks based on C/C++ languages,
Python, Hadoop using modern MapReduce computing method. Further
comparative studies concerned: adapting FITS data format to Hadoop
files management methods, combining Python and C++ for inter-process
communication in the context of the real-time data
streaming~\cite{andrea} with FPGA and GPU hardware accelerators.
\subsection{Archives}
The CTA Archives development are inspired by the lessons learned by
operating astronomical Observatory. The global architecture has been
designed based on the ISO OAIS Reference Model. Main prototyping
activities have been dedicated to explore the new database solutions
as a function of data type, volume and expected rate and type of
archive queries, e.g. the open-source, high-performance MongoDB
prototype both for housekeeping information and scientific raw
data.
\subsection{Observer Access}
Main prototyping activities are around the analysis software for open
access $ctools$: a set of analysis executables that is largely
inspired from HEASARC's ftools and Fermi's Science Tools, allowing the
assembly of modular workflows for CTA data analyses.The $ctools$ comprise
so far executables to simulate CTA event lists, to select and bin the
data, to perform a maximum likelihood analysis, and to create sky
images (see also~\cite{oa}).
\subsection{Information and Computing infrastructures}
The majority of the computing resources for the CTA massive MC
simulation production and analysis are linked to each other with the Grid
e-infrastructures and they have allowed DATA to evaluate the Grid
approach through a dedicated pathfinder initiative: CTACG. Among the
different prototyping activities within CTACG, the DIRAC (Distributed
Infrastructure with Remote Agent Control) WMS prototype was
intensively tested and evaluated as a solution to handle the MC
production. DIRAC was identified as useful and appropriate for the
following purposes: (i) Production job handling: e.g. for pipelines
(Monte-Carlo Production, Data reduction pipeline, On- site
Reconstruction and Analysis); (ii) User analysis job handling and data
management~\cite{dirac}.  Some prototyping activities were conducted
in order to build up an SAS through the web-base Science Gateway for
CTA. Two complementary solutions based on the same underlying portal
middleware $LifeRay$ were developed with different aims: the first one
integrating existing CTA Applications in a specific InSilicoLab
framework, the second one with a specific focus on Authentication,
Authorization and Single Sign On, with an architecture based on WS-
PGRADE/gUSE framework for integration of applications, and more
recently enriched with an interactive desktop environment named
ACID. A third prototype, compliant with the VO, has been developed
based on the Django framework. DATA is currently building a final
Gateway prototyping merging the complementary services provided by the
these three examples.
\section{Organization}
\begin{figure*}[!t]
 \center
 \includegraphics[width=11cm]{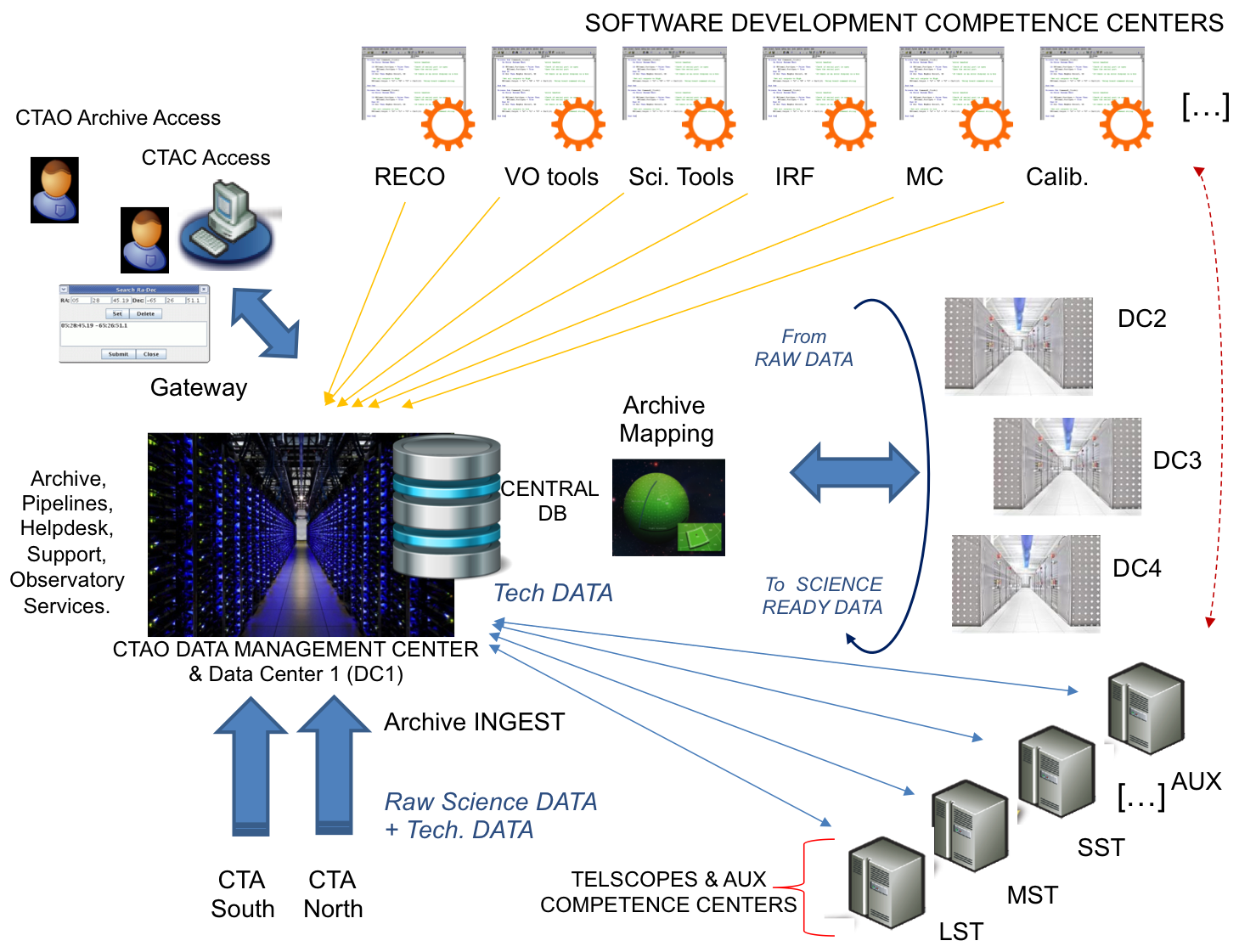}
 \caption{Logical diagram of the CTA Observatory functional Units. The CTAC competence centres guarantee the software and DATA services upgrading, while the CTAO will run them into the CTAO Data Management Centre. The ``telescopes and auxiliaries competence centres'' are those expert groups in any specific antenna, camera, device needing access to the Tech data (archived in DC1 and made available by the CTAO) for off-site monitoring purpose. The link with the Calibration competence centres will guarantee that all major changes in the software, which depend on Tech data are taken into account during the upgrading.}
 \label{fig:logicmodel}
\end{figure*}

The organization of the activities within DATA is structured around
the evolution of the overall CTA project, the complementary
of the CTAC and CTAO roles and their respective internal managements,
and the proposed computing and functional models. In Figure~\ref{fig:logicmodel} an overview of the logical implementation of the DATA baseline design is represented.
 At the same time some growing operational activities are managed by DATA: e.g. the software development services centrally managed at the computing centre CCIN2P3 and supported by a dedicated international ``CTA support group''; the DCI-Grid resources technical coordination for massive MC production (e.g. CTACG), including more than twenty computing centres distributed in several countries (France, Germany, Poland, Italy, Spain and others). During the construction phase, the DATA project
organization will evolve to
take into account the requirements settled on by the test and
pre-production operation phases of an increasing number of DATA products. The CTAC will organize itself in ``competence centres'' such as specific and continuous software development groups or technical and auxiliary devices monitoring groups (see Figure~\ref{fig:logicmodel}). In the construction phase the same competence groups will be in charge of bringing in operation and support their products. During the operation phase some key experts within the CTAO will have
the commitment to guarantee the operation and maintenance of any piece
of software and DATA services, while the competence centres will be
the CTAC instances, which will guarantee the software and services
upgrading according to the contingent needs. The scientific and operative link between CTAC and CTAO in DATA will be represented by the shared scientific analysis platform.   

\section{Conclusions}
A technical design of the complete data life cycle and management for the CTA observatory has been finalized. A set of solutions and a series of prototypes have been proposed to organize the Science Data Center of the CTA observatory. In the next years, entering in the CTA construction phase, an intense implementation activity is expected within the DATA project. Some major technical choices will be required and the computing model could also evolve in consideration of the CTA sites final location, new requirements and the potential evolution of computing architectures. \acknowledgments
We gratefully acknowledge support from the agencies and organisations listed at the following URL: http://www.cta- observatory.org/?q=node/22

\end{document}